\documentclass{aa}  
\usepackage{graphicx}
\usepackage{txfonts}
\usepackage{natbib}
\newcommand{\HeI}{\ion{He}{i}}

\newcommand{\HI}{\ion{H}{i}}
\newcommand{\HII}{\ion{H}{ii}}
\newcommand{\OIII}{[\ion{O}{iii}]}

\newcommand{\ArII}{[\ion{Ar}{ii}]}
\newcommand{\ArIII}{[\ion{Ar}{iii}]}
\newcommand{\SIII}{[\ion{S}{iii}]}
\newcommand{\SIV}{[\ion{S}{iv}]}
\newcommand{\NeII}{[\ion{Ne}{ii}]}
\newcommand{\NeIII}{[\ion{Ne}{iii}]}

\newcommand{\um}{$\mu$m}

\newcommand{\spitzer}{{\it Spitzer}}

\begin{document}
   \title{A mid-infrared study of \HII{} regions in the Magellanic 
          Clouds: \\ N88\,A and N160\,A
          }

   \author{N.\,L. Mart\'{\i}n-Hern\'{a}ndez
          \inst{1}
          \and
           E. Peeters\inst{2,3,4}
          \and
          A.\,G.\,G.\,M. Tielens\inst{5}
          }

   \offprints{N.\,L.\,Mart\'{\i}n-Hern\'{a}ndez}

   \institute{Instituto de Astronom\'ia de Canarias, V\'ia L\'actea s/n,
             E\,38205 La Laguna, Spain\\
              \email{leticia@iac.es}
         \and
         NASA Ames Research Center, MS 245-6, Moffett Field, CA 94035, USA
         \and
	 SETI Institute, 515 N. Whisman Road, Mountain View, CA 94043, USA
	 \and
	 Department of Physics and Astronomy, The University of Western Ontario, London, 
	 ON N6A 3K7,  Canada
         \and
         NASA Ames Research Center, MS 245-3, Moffett Field, CA 94035, USA
	 }

   \date{Received; accepted}

 
  \abstract
  {}  
   {To show the importance of high-spatial resolution observations of \HII\ regions when compared with observations obtained with larger apertures such as ISO, we present mid-infrared spectra of two Magellanic Cloud \HII\ regions, N88\,A and N160\,A. }
   {We obtained mid-infrared ($8-13$\,\um), long-slit spectra with TIMMI2 on the ESO 3.6 m telescope. These are combined with archival spectra obtained with the Infrared Spectrograph (IRS) onboard the \spitzer\ Space Telescope, and are compared with the low-spatial resolution ISO-SWS data. An inventory of the spectra in terms of atomic fine-structure lines and molecular bands is presented.}
   {Concerning N88\,A, an isolated \HII\ region with no adjacent infrared sources, the observations indicate that the line fluxes observed by ISO-SWS and \spitzer-IRS come exclusively from the compact \HII\ region of about 3\arcsec\ in diameter.  This is not the case for N160\,A, which has a more complex morphology. We have spectroscopically
   isolated for the first time the individual contributions of the three components of N160\,A, two high-excitation blobs, A1 and A2, and the young stellar object N160\,A-IR.  In addition, extended \SIV\ emission is observed with TIMMI2 and is most likely associated with the central star cluster located between A1 and A2. We show the value of these high-spatial resolution data in determining source characteristics, such as the degree of ionization of each high-excitation blob or the bolometric luminosity of the YSO. This luminosity ($2\times10^5\,L_\odot$) is about one order of magnitude lower than previously estimated. For each high-excitation blob, we also determine the electron density and the elemental abundances of Ne, S, and Ar.}
   {}


\titlerunning{Mid-infrared study of the \HII{} regions N88\,A and N160\,A}

   \maketitle
%

\section{Introduction}

High-Excitation Blobs (HEBs) are compact \HII\ regions that constitute a rare class of ionized nebulae in the Magellanic Clouds \cite[e.g.][and references therein]{heydari07}. These objects are tightly linked to the early stages of massive star formation, when the stars disrupt their natal molecular clouds. They are considered to be the first step in the evolution from an ultracompact \HII\ region towards more extended structures. Their main characteristics are: high excitation, small size \hbox{($\sim2$ pc)}, high density, and higher extinction than the typical \HII\ regions in the Magellanic Clouds. So far, only 6 members have been discovered in the Large Magellanic Cloud (LMC) and 3 in the Small Magellanic Cloud (SMC). Most of these HEBs have a counterpart observed in radio \citep{indebetouw04,martin:atca:mc}.

Infrared spectroscopy is one of the best ways to determine the characteristics of \HII\ regions.
The overall mid-infrared (MIR) spectra of \HII\ regions \citep[e.g.][]{peeters:catalogue} are dominated by atomic fine-structure lines that originate in the ionized gas, a strong rising continuum due to thermal
emission of dust and strong, broad emission features at 3.3, 6.2, 7.7,
8.6, 11.2 and 12.7 \um, attributed to emission by
Polycyclic Aromatic Hydrocarbon (PAH) molecules \citep[e.g][]{allamandola89}.

The comparison of ISO-SWS spectra of \HII\ regions in the Magellanic Clouds (which also include a few HEBs) and in our Galaxy shows a clear segregation in PAH ratios between the sources of different type of environment, i.e. from the Milky Way to the LMC and to the SMC \citep{vermeij02}. This is most certainly due to the different environment encountered in the Magellanic Clouds, in particular, its low metal content. This study, nevertheless, is based upon observations made with the large ISO aperture (between 14\arcsec\ and 33\arcsec), where other contributions besides that of the \HII\ region itself are likely to exist.

\begin{table*}
\caption{Journal of the TIMMI2 observations.}
\label{tbl:log}
\begin{center}
\begin{tabular}{l c c c c c c} 
\hline\hline\\[-5pt]
Source & RA ($^h$ $^m$ $^s$) & Dec ($\degr$ $\arcmin$ $\arcsec$) & Date & Standard & \multicolumn{2}{c}{Flux (mJy)}\\ \cline{6-7}
& \multicolumn{2}{c}{(J2000.0)} & & star & N1 & N10.4\\[2pt]
\hline\\[-5pt]

N88A     & 01 24 07.9  & $-$73 09 04 & 2003-11-14 & HD4815
&$100\pm40^{\rm (a)}$&--\\
N160\,A1 & 05 39 43.3  & $-$69 38 54 & 2003-11-17 & HD39523 &--&
$1330\pm270^{\rm (b)}$\\
N160\,A2 & 05 39 45.9  & $-$69 38 39 & 2003-11-17 & HD39523 &--&
$1630\pm330^{\rm (b)}$\\
N160A-IR & 05 39 43.9  & $-$69 38 34 & 2003-11-14 & HD39523 &--& $<40^{\rm (c)}$\\

\hline
\end{tabular}
\end{center}
$^{\rm (a)}$ The image was calibrated using the spectrum since we did
not image the standard star in this filter. Slit losses of
up to 50\% are considered in the uncertainty. 
$^{\rm (b)}$ The uncertainty ($\sim20$\%) accounts for variations in the transparency of
the atmosphere and the absolute flux calibration error of the standard star.
$^{\rm (c)}$ Upper limits are defined as the flux of a point source with
a peak flux three times the image noise.
\end{table*}

\begin{table}[!tp]
\caption[The IRS sample]{Journal of the \spitzer\ observations. The 
coordinates of the positions observed with \spitzer-IRS are given. }
\label{tbl:spitzer}
  \begin{center}
    \leavevmode
\begin{tabular}{l@{\hspace{3pt}}c@{\hspace{1pt}}c@{\hspace{5pt}}c@{\hspace{5pt}}c}
\hline\hline\\[-5pt]
Source & 
\multicolumn{1}{c}{RA ($^h$ $^m$ $^s$)} &
\multicolumn{1}{c}{Dec ($\degr$ $\arcmin$ $\arcsec$)} & 
\multicolumn{2}{c}{AORKEY$^{\rm (a)}$} \\
&\multicolumn{2}{c}{(J2000.0)}  &\\[5.pt]
\hline \\[-5.pt]

& \multicolumn{2}{c}{IRAC}\\

N88\,A     & & & 4383488 \\

N160\,A    & & & 12546816\\[5pt]

\hline \\[-5.pt]

& \multicolumn{2}{c}{IRS}  & \multicolumn{1}{c}{SH/LH} & \multicolumn{1}{c}{SL/LL}\\

N88\,A     & 1 24 08.10 & $-$73 09 05.00 &  4384000 \\

N160\,A2   & 5 39 46.10 & $-$69 38 36.00 & 12606208 & 8242944\\
           & $-$ & $-$ & 12902556 & 8568832\\
           & $-$ & $-$ & 13048576 & 8579328\\
           & $-$ & $-$ & 13162752 & 11812352\\
           & $-$ & $-$ & 13358336 & 12003584\\
           & $-$ & $-$ & 13529088 & 12130048\\
           & $-$ & $-$ & 15672064 & 15672064\\
           & $-$ & $-$ & 16295936 & 16295936\\
           & $-$ & $-$ & 16702976 & 16702976\\

\hline \\[-5.pt]

& \multicolumn{2}{c}{MIPS}\\
N160\,A  & & & 14381312 \\

\hline \\[-0.5cm]
\end{tabular}
  \end{center}
$^{\rm (a)}$ AORKEY which uniquely identifies each \spitzer\ observation.
\end{table}

MIR spectra of \HII\ regions give also access to various elements, e.g.
Ar, S and Ne, in two different ionization stages. Ratios of
fine-structure lines such as \ArIII/\ArII\ 9.0/7.0, \SIV/\SIII\
10.5/18.7 and \NeIII/\NeII\ 15.5/12.8 \um\ probe the ionizing stellar
spectrum between 27.6 and 41 eV and are widely used to constrain the
properties of the ionizing stars in \HII\ regions.  Again, the use of
large apertures is limited since the contributions from neighboring sources cannot be disentangled.

High-spatial resolution observations are thus essential to provide
accurate measurements of the line (and PAH) fluxes emitted by the
object itself.
In this sense, the study of \HII\ regions in the Magellanic Clouds, and specially of HEBs due to their compactness, might serve as a stepping stone to understand \HII\ regions in other galaxies, where their emission is confined into one beam. \HII\ regions in the Magellanic Clouds are close enough that we can still resolve their individual components. Hence, we can disentangle their contribution to the overall spectrum and assess how much the properties of the \HII\ region depend on aperture size. This is actually difficult to do for \HII\ regions in our Galaxy where the surrounding ionized gas, because of its low density, is usually too diffuse and extended to be well sampled.

In this paper, we study in detail the spectral morphology of two well distinct regions, N88\,A and N160\,A, at high-spatial resolution using TIMMI2 and the Infrared Spectrograph (IRS) onboard the \spitzer\ Space Telescope. These spectra are subsequently compared with findings (and conclusions) based upon large aperture data obtained with ISO-SWS.
N88\,A, located in the SMC, is a compact HEB which is among the brightest objects in the Magellanic Clouds \citep[e.g.][]{heydari99c}. N160\,A, located in the LMC, is a particularly interesting region since it harbors two HEBs named A1 and A2 \citep[][]{heydari86} and a  massive Young Stellar Object \citep[YSO;][]{epchtein84,henning98},  called N160\,A-IR. For N160\,A, we have {\it spectroscopically} isolated for the first time the individual contributions of the two HEBs and the YSO. The images in the near- and mid-infrared of \cite{henning98} clearly separate this YSO from the neighboring sources, but the spectrum published by these authors, obtained with ISO, is dominated by emission from the surrounding ionized gas. 

The core of this paper is structured as follows. Section~\ref{sect:obs} presents the observations and data reduction. Section~\ref{sect:desc} describes the general morphology of N88\,A and N160\,A. Results are presented in Sect.~\ref{sect:results}, which includes a description of the MIR spectra, a comparison between observations obtained with different apertures, an estimate of the physical conditions of the ionized gas (in terms of electron density and elemental abundances) and, finally, the modeling of the Spectral Energy Distribution (SED) of the YSO in N160\,A. The main conclusions are summarized in Sect.~\ref{conclusions}.

\begin{table}[!tp]
\caption[The sample]{Journal of the ISO observations. The coordinates of the
  positions observed with ISO-SWS are given. }
\label{tbl:sws}
  \begin{center}
    \leavevmode
\begin{tabular}{l@{\hspace{1pt}}c@{\,}c@{\hspace{3pt}}c@{\hspace{4pt}}c@{\hspace{4pt}}l}
\hline\hline\\[-5pt]
Source & 
\multicolumn{1}{c}{RA ($^h$ $^m$ $^s$)} &
\multicolumn{1}{c}{Dec ($\degr$ $\arcmin$ $\arcsec$)} & 
\multicolumn{1}{c}{Date} & 
\multicolumn{1}{c}{TDT$^{\rm (a)}$} & 
AOT\\
&\multicolumn{2}{c}{(J2000.0)} & &\\[5.pt]
\hline \\[-5.pt]

N88\,A     & 01 24 08.15 & $-$73 09 03.2 & 1997-09-08 & 66300110 & 01$^{\rm (b)}$ \\

N160\,A2   & 05 39 46.12 & $-$69 38 36.6 & 1997-08-05 & 62804104 & 01$^{\rm (b)}$\\
N160\,A-IR & 05 39 43.74 & $-$69 38 30.4 & 1997-08-01 & 62401303 & 01$^{\rm (c)}$\\


N160\,A1   & 05 39 43.27 & $-$69 38 51.4 & 1996-07-10 & 23600925 & 02 \\[5pt]

\hline \\[-0.5cm]
\end{tabular}
  \end{center}
$^{\rm (a)}$ Target Dedicated Time (TDT) which uniquely identifies each
ISO observation.
$^{\rm (b)}$ Speed 2. 
$^{\rm (c)}$ Speed 3. 
\end{table}

\section{Observations and data reduction}
\label{sect:obs}
\subsection{TIMMI2}
\label{timmi2:obs}

\begin{figure*}
   \centering \resizebox{\hsize}{!}{
   \includegraphics{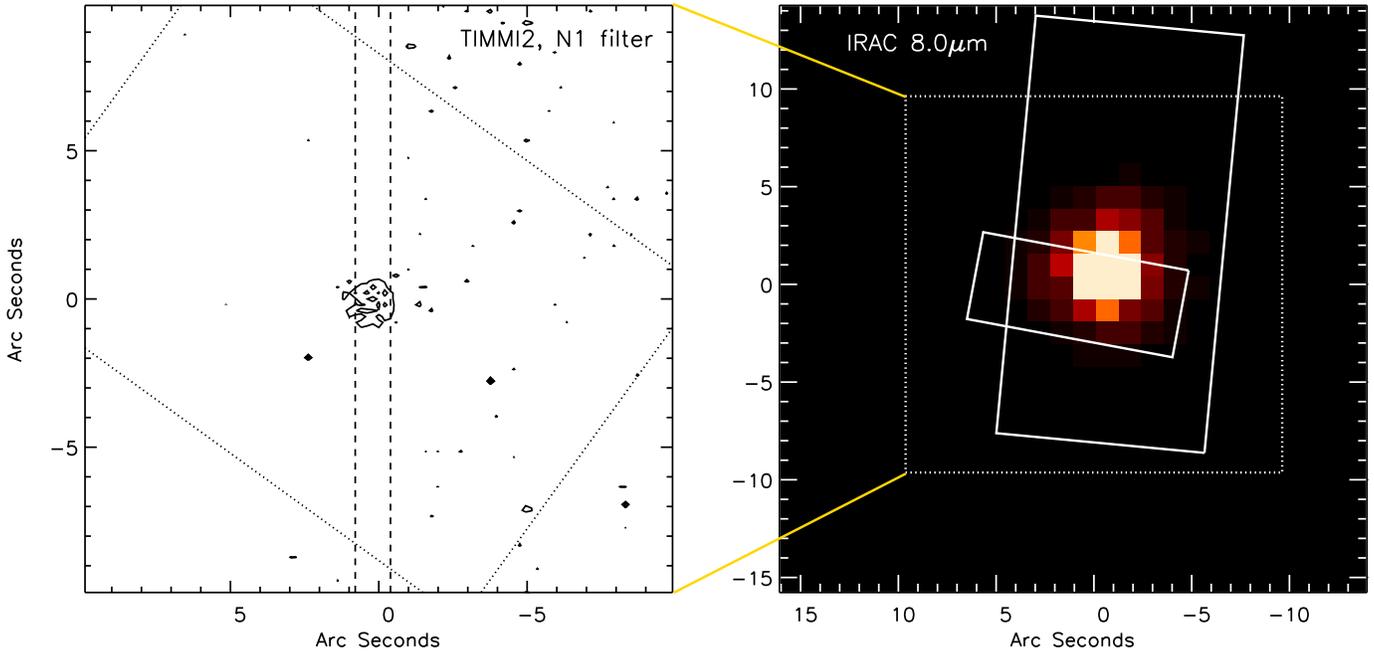}}
     \caption{{\it (Left)} Contours of the N1  ($\lambda_{\rm c}=8.70\,\mu$m) image of N88\,A. 
        Contour levels are 2.5, 4 and 6 times 
        the image rms  ($\sigma=0.05$ mJy/pixel). 
	The ISO-SWS (band 2C, between 7.0\,\um\ and 12.1\,\um) aperture (dotted lines) and the
        TIMMI2 slit (dashed lines) are indicated.
	{\it (Right)} \spitzer-IRAC image of N88\,A at 8.0~\um.
	The \spitzer-IRS SH and LH apertures (only nod 1) are plotted by solid lines. The dotted box outlines the field-of-view covered by the left panel.
        North is up and East is to the left. 
        The (0\arcsec, 0\arcsec) position in both panels corresponds to 
        $\alpha$(J2000)=$1^{\rm h} 24^{\rm m} 7\fs920$ and
        $\delta$(J2000)=$-73{\degr} 9{\arcmin} 4\fs07$
        [see electronic edition of the Journal for a color version of this figure].
        }         
         \label{fig:slits}
   \end{figure*}

The $N$-band imaging and spectroscopic observations were made with
the Thermal Infrared MultiMode Instrument (TIMMI2) at the ESO 3.6\,m telescope. A journal of these observations is presented in Table~\ref{tbl:log}.
The camera uses a $320 \times 240$ pixel Raytheon Si:As array. In imaging, we
selected the 0\farcs2 pixel scale providing a total field of view of 
$64\arcsec \times 48\arcsec$. For long-slit spectroscopy, 
we used the 10~\um\ low-resolution grism which ranges from 
7.5 to 13.9~\um\ and has a spectral resolving power $\lambda/\Delta\lambda \sim 160$. The slit
used was 1\farcs2$\times$70\arcsec, with a pixel scale of 0\farcs45.
It was oriented along P.A.$=0$\degr\ covering the brightest nebular spots of N88\,A 
(cf. Fig.~\ref{fig:slits}) and N160\,A  (cf. Fig.~\ref{fig:rgb}). The three slits covering N160\,A are denominated, from west to east, s1, s2 and s3. They were selected to cover, respectively, N160\,A1, N160\,A-IR and N160\,A2. The infrared seeing during our observations varied between 1\farcs6 and 1\farcs9, measured from the full-width-at-half-maximum (FWHM) of the standard star spectra.

In order to correct for background emission from
the sky, the observations were performed using a standard
chopping/nodding technique. 
For N88\,A, we used chop and nod amplitudes of 15\arcsec\ for imaging, and of
10\arcsec\ for spectroscopy. To image N160\,A, we used an amplitude of 30\arcsec;
for spectroscopy, we used 20\arcsec\ in the case of slit s3, and 
30\arcsec\ in the case of slits s1 and s2.  

The data processing included the removal of bad frames and the co-addition
of all chopping and nodding pairs. This procedure is applied to both the 
target and the calibration standard star. We propagated the uncertainty 
for each pixel, dominated by variations of the sky transparency, along 
each step of the processing.

The $N$-band imaging of the standard stars
was performed in the same filter as
the target and used for photometric flux conversion from photon count
rates (ADU/s) into astronomical units (Jy).
The MIR flux densities of the targets in the resulting images are 
given in Table~\ref{tbl:log}. N88\,A was observed in the N1 filter (red and blue cuts are 7.79 and 9.46\,\um, respectively), and N160\,A in the N10.4 filter (9.46 and 11.21\,\um, respectively).
Astrometry was performed by 
comparing with radio maps 
\citep[cf.][]{indebetouw04,martin:atca:mc} and by matching the positions of stars in common in our images and in the 2MASS survey catalogue. The final (contour) images are shown in the left panels of Figs.~\ref{fig:slits} and \ref{fig:rgb}.

The spectra of the standard stars were used for atmospheric
corrections and for flux calibration.
To minimize residuals of the sky line correction, we observed
the standard star and the target at similar airmasses. In addition, we
applied a flux correction provided by the TIMMI2 instrument team that
takes into account the differential extinction between the target and
the standard star.
The synthetic calibrated spectra for these standard stars are given by
\citet{cohen99}.
We note that, close to the blue and red cut-off frequencies of the grating, 
the terrestrial background varies
strongly with ambient conditions and atmospheric corrections are more
difficult. The same holds for wavelength regions of strong terrestrial
lines near 9.58\,\um, 11.73\,\um\ and 12.55\,\um.
Finally, wavelength calibration is straight forward since a
table with the pixel-to-wavelength correspondence is provided by the
TIMMI2 team.

Spectra were extracted using apertures of 1.5 
times the FWHM measured along the 
spatial direction, where 
FWHM(N88A)\,= 3\farcs0, 
FWHM(N160\,A1)\,= 3\farcs1,  
FWHM(N160\,A2)\,= 5\farcs3  and
FWHM(N160A-IR)\,= 2\farcs3.
The resulting spectra are shown in Figs.~\ref{fig:n88:sp},
\ref{fig:n160:irs} and .\ref{fig:n160:sp} 

\subsection{Spitzer}
\label{spitzer:obs}

N88\,A and N160\,A were also observed with the IRAC instrument
\citep{fazio04}, the MIPS instrument \citep{rieke04}, 
and the IRS instrument \citep{houck04} onboard the
\spitzer\ Space Telescope \citep{werner04}.  From the
\spitzer\ archive, we obtained the IRAC pbcd data associated with
pipeline version S14, the MIPS pbcd data associated with pipeline version S16, and the IRS Short-High (SH), Long-High (LH),
Short-Low (SL) and Long-Low (LL) bcd data associated with pipeline
version S15 (see Table~\ref{tbl:spitzer}). 
Some observations, although available, were not considered here. The LL spectrum of
N160\,A2, for instance, is not used since the LL wavelength region is
covered already by the SH/LH modes, which also have a higher spatial resolution. In addition, the LL (1$^{\rm st}$ order) spectrum of N160\,A-IR and the 24\,\um\ MIPS map of N160\,A were not considered due to saturation problems.

   \begin{figure*}
   \centering \resizebox{\hsize}{!}{
   \includegraphics{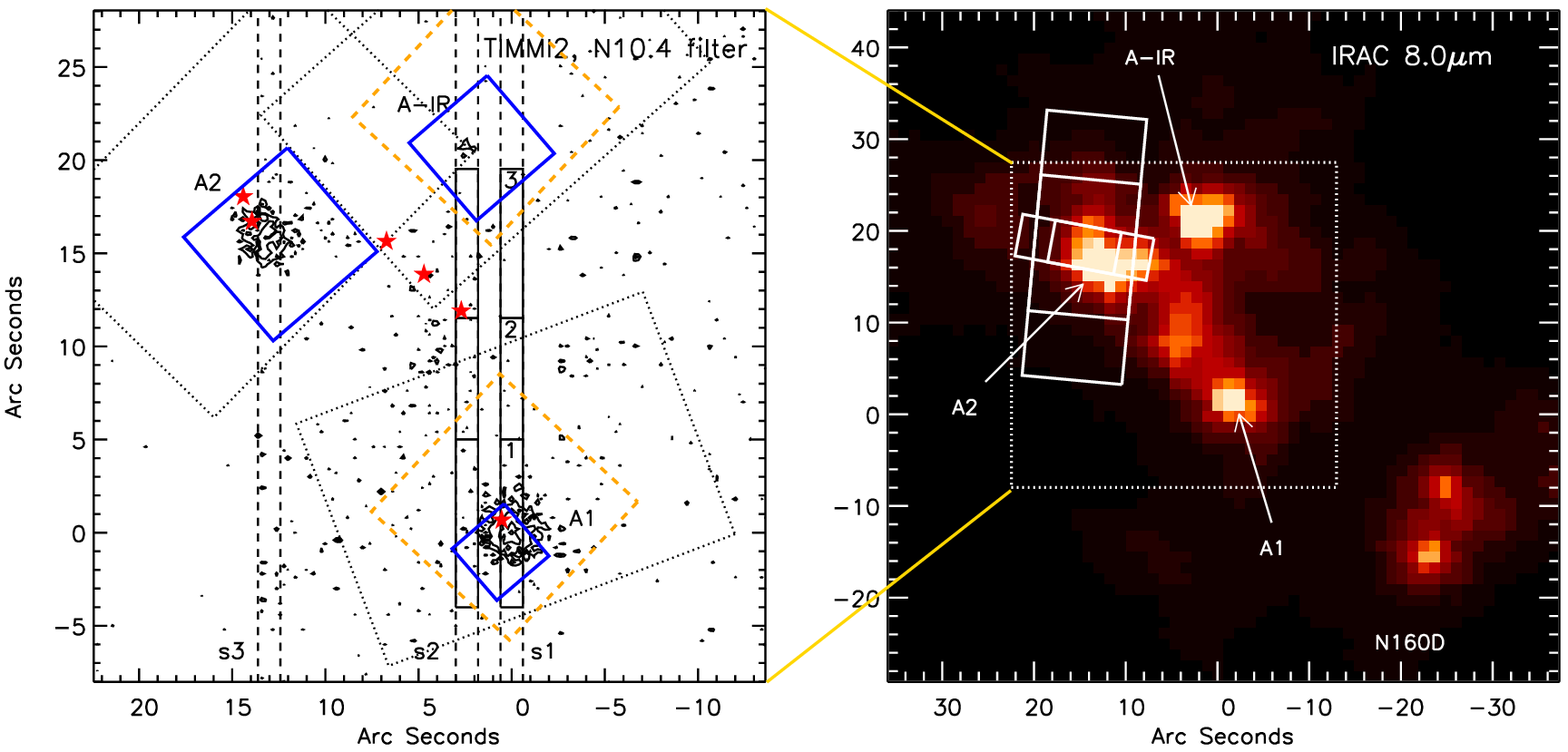}}
   \caption{
        {\it (Left)} Contours of the TIMMI2 N10.4 ($\lambda_{\rm c}=10.38\,\mu$m) image of N160A. 
        Contour levels are 2.5, 4, 7 and 
        10 times the image rms ($\sigma=0.21$ mJy/pixel).
        The red stars \citep[cf.][]{heydari02b} show the positions of 
        the brightest stars ionizing N160\,A1 (star \#22) and N160\,A2 
        (stars \#100 and \#105), and of 
        the brightest stars grouped in the central cluster (stars \#24,
        \#35 and \#58).
        The ISO-SWS (band 2C, between 7.0\,\um\ and 12.1\,\um) apertures 
        (dotted lines), the three TIMMI2 slits (dashed lines), 
	and the \spitzer-IRS SL (blue solid lines) and LL (orange dashed lines) apertures centered 
        on each component are 
        indicated.
        The TIMMI2 zones defined in Fig.~\ref{fig:n160a:siv}  (1, 2 and 3) 
        are also indicated (black solid lines).
        {\it (Right)} \spitzer-IRAC image of N160\,A at 8.0~\um.
	One of the \spitzer-IRS SH and LH apertures (solid lines, which correspond
	to AORKEY=15672064) is plotted as a reference.
        The dotted box outlines the field-of-view covered by the left panel.
	North is up and East is to the left.
        The (0\arcsec, 0\arcsec) position corresponds to 
        $\alpha$(J2000)=$5^{\rm h} 39^{\rm m} 43\fs291$ and
        $\delta$(J2000)=$-69{\degr} 38{\arcmin} 54\fs28$
        [see electronic edition of the Journal for a color version of this figure].}
  \label{fig:rgb}
   \end{figure*}

IRAC images of N88\,A and N160\,A are shown in the right panels of Figs.~\ref{fig:slits} and \ref{fig:rgb}. The IRS SH and LH apertures centered on N88\,A (only nod~1)
are indicated in the right panel of Fig.~\ref{fig:slits}. Here,
only the spectrum of nod\,1 is used because of the better overlap between the different orders. 
N160\,A2 was used as a
calibrator for IRS and therefore, it was observed multiple times using the same
pointing (see Table~\ref{tbl:spitzer}). The orientation of
each slit depends on the observing date/time.
The right panel of Fig. ~\ref{fig:rgb}, for instance, shows the slit of one specific SH and LH
observation centered on N160\,A2, i.e. AORKEY = 15672064. The other slits are rotated with respect to this one. 
IRS SL and LL spectra of N160\,A2 were also obtained through multiple slits. This and the large slit size allowed us to create spectral maps using CUBISM --see e.g. Fig.~\ref{fig:n160a:map}, where a \SIV\ 10.5\,\um\ line flux map has been created using this software-- and, consequently, to extract a spectrum for each individual component. These extraction apertures are shown in
the left panel of Fig.~\ref{fig:rgb}. Full-slit aperture extraction was performed for the SH
and LH observations.
The aperture sizes are listed in Table~\ref{tbl:fluxes}.

Additional processing was applied to the IRS bcd data. Rogue pixels
and bad pixels identified in the bmask were replaced with
IRSCLEAN. This program obtains the new flux values by deriving
profiles for a group of rows above and below the bad pixel
(i.e. neighboring wavelengths), predicting the profile for the bad
pixel's row, and scaling that profile by the good pixels in the bad
pixel's row. Residual bad data
were manually removed from the obtained spectra. Note that background
subtraction has not been applied.
The resulting SH and LH spectra
have a resolving power of 600. The SL and LL spectra, on the other hand, have a resolving
power ranging from 60 to 128. They are shown in Figs.~\ref{fig:n88:sp}, \ref{fig:n160:irs} and \ref{fig:n160:sp}.

Given that N160\,A2 is extended, different slit orientations result
in flux level variations up to 20\% in the resulting individual SH and LH spectra (because of the full-slit aperture extraction).  
For the final line fluxes reported in Table~\ref{tbl:fluxes},
we consider the mean of the fluxes in all positions. Uncertainties take into account these variations from slit to slit.

In addition, in {\it all} SH 2-dimensional (2d) bcd images of N160\,A2,
there are between 1 and 4 pixels in the
\NeIII\ 15.5\,\um\ fine-structure line that only have  1 useable
(i.e. non-saturated) plane in the raw data making it impossible to
obtain a reliable flux for the considered pixel. These (saturated)
pixels are corrected for by IRSCLEAN, which operates on the 2d bcd
images. However, the accuracy of the corrected flux decreases with
the increasing number of bad pixels. Nevertheless, when comparing
the \NeIII\ 15.5\,\um\ line flux of N160\,A2 for all SH observations,
no correlation is found with the amount of pixels that have 1 single
useable plane, and all line fluxes fall within $\pm 10\sigma$ of the mean
value, similar to the range in fluxes found for the other fine-structure lines. This suggests that the variations we see in the \NeIII\ line flux are most likely due to the different slit orientations (as mentioned previously).
The \NeIII\ 15.5\,\um\ line flux should anyway be considered with caution.

\subsection{ISO-SWS}

N88\,A and N160\,A were observed with the Short Wavelength Spectrometer
\citep[SWS,][]{deGraauw96} onboard the Infrared Space
Observatory \citep[ISO,][]{kessler96} in the AOT~01 and 
AOT~02 modes
with a resolving power ($\lambda/\Delta\lambda$) ranging from 350 to 1500 
(see Table \ref{tbl:sws}). 
The AOT~01 mode covers the full SWS wavelength range from 2.3 to 45\,\um, 
while the AOT~02 mode consists of separate line scans.
The AOT~01 data were processed
with IA$^3$, the SWS Interactive Analysis package using calibration
files and procedures equivalent with pipeline version 10.1.  
A detailed account of the reduction can be found in
\citet{peeters:catalogue}. For the AOT~02 observations, we refer to
\citet{vermeij:data} and \citet{vermeij:analysis}.
The ISO spectra of N160\,A2 and the YSO N160\,A-IR are plotted in Fig.~\ref{fig:n160:sp}.
Note that the ISO-SWS observation centered on A1 was made in AOT~02 mode
and consists only of separate line scans;  hence, this spectrum is not shown.

\section{Source characteristics}
\label{sect:desc}

\subsection{N88\,A}
\label{desc:n88a}

N88\,A is among the brightest objects in the Magellanic Clouds and is
part of the large \HII\ complex N88   (\citealt{henize56}; other designation is DEM\,161, \citealt{davies76}), 
located east of the main body of 
the SMC. It is a compact (about 3\arcsec\ in diameter) and high excitation \HII\ region 
\citep[cf.][]{heydari99c}, unusual among SMC \HII\ regions because
it contains much dust \citep[e.g.][]{testor85,kurt99}. The interstellar
reddening towards N88\,A,  as derived from the Balmer decrement in the visible, is on average $A_{\rm V}\sim 1.5$~mag 
\citep[cf.][]{heydari99c,testor05}, unexpectedly high for an \HII\
region in the metal poor SMC. Hence, \cite{heydari99c} describe
N88\,A as a HEB, a very young \HII\ region which is hatching from its natal 
molecular cloud and is heavily affected by absorbing dust associated
with the cloud.

N88\,A has a complex and inhomogeneous morphology \cite[cf.][]{heydari99c}.
An absorption lane crosses the nebula from north to south and here 
the extinction rises to more than 3.5~mag. West of this dusty 
structure lies the brightest part of N88\,A, a small core of 0\farcs3 in
diameter. A sharp edge is visible on the north-west and several
lower excitation arc-shaped features and filaments seem to emerge from the
main body of the source.
Its radio continuum emission  \citep[cf.][]{indebetouw04} agrees fairly
well with the global optical morphology, although it appears to be 
slightly offset to the west of the H$\alpha$ emission. These authors
note, however, that this offset could be due to the $\sim 1$\arcsec\ 
pointing uncertainty of HST. 

\cite{heydari99c} also show that the bright H$\alpha$ core of N88\,A contains 
at least two stars (\#1 and \#2), and a third fainter one is located 
to the east, marginally detected and thus more embedded. 
These authors suggest that this faint star might be the exciting star.
\cite{testor03} derived a spectral type of O6--O8\,V for the ionizing star. This spectral type classification is strengthened by the nebular near-infrared spectrum obtained by \cite{testor05}, where the ratio of the 2.11\,\um\ \HeI\ to Br$\gamma$ is consistent with an ionizing source of spectral type earlier than O7\,V \citep[see table 3 of][]{hanson02}.

Figure~\ref{fig:slits} shows the TIMMI2 (N1 filter) and \spitzer-IRAC (8~\um) images of N88\,A. On top, we show the TIMMI2 slit, and the \spitzer-IRS\ and ISO-SWS larger apertures.

\subsection{N160\,A}
\label{desc:n160a}

N160  (\citealt{henize56}; other designations are DEM\,284, \citealt{davies76}, 
and MC\,76, \citealt{mcgee66}) is a giant \HII\ complex lying about 30\arcmin\
south of 30\,Doradus. The complex is
formed by two main regions $\sim$3\farcm5 apart
\citep{heydari86}. The brighter western region comprises components A
(NGC\,2080) and D (NGC\,2077), while the eastern region is formed by
components B (NGC\,2085) and C (IC\,2145). N160 is linked with the OB
star association LH\,103 \citep{lucke70}.

N160A is the brightest component of N160. It is a particularly
interesting region since it harbors two HEBs
\citep[called A1 and A2, and discovered by][]{heydari86} and a young
stellar object (named N160\,A-IR). This YSO was discovered using near- and mid-infrared photometry by
\cite{epchtein84} and has been further observed in
the infrared by \cite{henning98}. 
A water
maser has been found at the interface between the molecular cloud and the \HII\
region providing strong evidence that feedback from massive stars is triggering subsequent star formation \citep{oliveira06}. 

HST observations of N160\,A \citep{heydari02b} reveal a bright nebula of about 35\arcsec\
by 25\arcsec. The compact \HII\ region
A1, which is the brightest part of N160\,A, shows a tiny cavity of some
2\farcs3 across which might be the result of the strong wind of a
bright star found towards the center of the blob. This star must be a
massive O type since it has carved the cavity and produces the
highest \OIII/H$\beta$ ratio in the whole region. The other HEB, A2,
is 3\arcsec\ in diameter, has a patchy appearance marked by the
presence of several thin absorption lanes and contains about dozen
rather faint stars. Three bright cluster stars are also identified within the
nebula along with the two blobs A1 and A2. 
These three stars seem to contribute to the ionization of N160\,A. The
visual magnitude of the brightest member of this central cluster agrees
with an O6.5V star.

The A1 and A2 blobs lie in very dust rich areas of the region. In the case of A1, dust 
is mainly located behind the southern border of the cavity where the mean value of the 
H$\alpha$/H$\beta$ Balmer decrement gives an extinction $A_{\rm V}=1.3$~mag. The
northern compact \HII\ region A2 is more affected by dust, with a mean extinction 
$A_{\rm V}=1.5$~mag \citep[cf.][]{heydari02b}.

The radio continuum emission observed towards N160\,A reveals an extended
structure coincident with the optical nebula where the two HEBs A1 and
A2 stand out as two strong radio peaks \citep[cf.][]{indebetouw04,martin:atca:mc}. As
expected from its youth, the YSO does not have associated radio
continuum emission.

Figure~\ref{fig:rgb} shows the TIMMI2 (N10.4 filter) and \spitzer-IRAC (8\,\um) images of N160\,A. The YSO is not registered by the TIMMI2 image but it appears as a very bright source in the IRAC one. The larger field of view of the \spitzer-IRAC image also registers the western N160\,D region. The TIMMI2 slits, and the \spitzer-IRS\ and ISO-SWS apertures are indicated.


\section{Results}
\label{sect:results}

\subsection{The MIR spectrum}

We present here the high-spatial resolution spectroscopy data obtained with TIMMI2 and \spitzer, which will be compared with ISO observations. 

A variety of fine-structure lines and broad-band dust features fall within the ground-based $N$-band spectroscopic range \hbox{(8--13 \um)} of TIMMI2. The most relevant lines are \ArIII\ at 9.0\,\um, \SIV\ at 10.5\,\um\ and \NeII\ at 12.8\,\um. These lines require hard radiation with energies between $\sim 21$ and 35~eV, and the most likely explanation for their excitation mechanism is photoionization by hot stars. \spitzer-IRS\ and ISO-SWS cover a wider wavelength range, including lines such as \ArII\ at 7\,\um, \NeIII\ at 15.5 and 36\,\um, and \SIII\ at 18.7 and 33.5\,\um.

  \begin{figure}
   \centering \resizebox{\hsize}{!}{
     \includegraphics{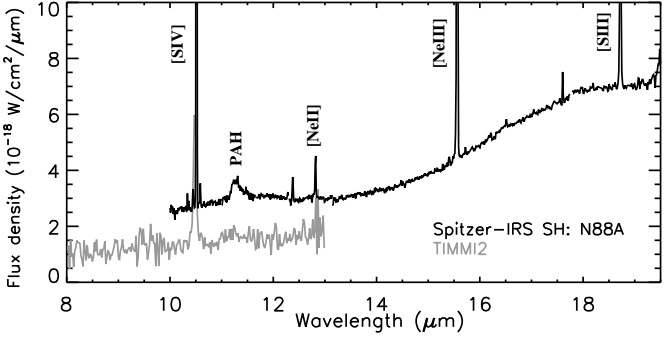}}
     \caption{\spitzer-IRS SH spectrum of N88A. The TIMMI2 spectrum is plotted in grey.  The most important atomic lines and PAH bands are labeled. Note that this \spitzer\ spectrum is not scaled down.}
        \label{fig:n88:sp}
   \end{figure}
 
  \begin{figure}[!h]
   \centering \resizebox{\hsize}{!}{
   \includegraphics{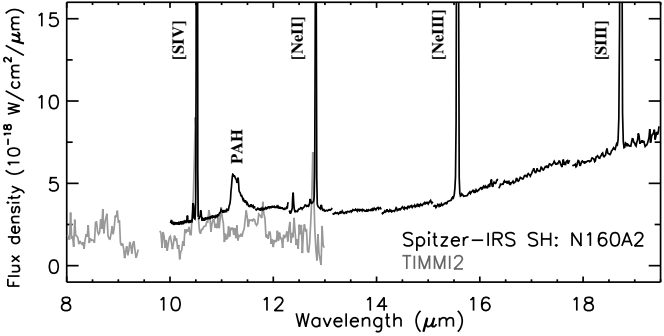}}
      \caption{\spitzer-IRS SH spectrum of N160\,A2. The TIMMI2 spectrum is plotted in grey. Note that the \spitzer-IRS\ spectrum is not scaled down.}
         \label{fig:n160:irs}
   \end{figure}

  \begin{figure}[!ht]
   \centering \resizebox{\hsize}{!}{
   \includegraphics{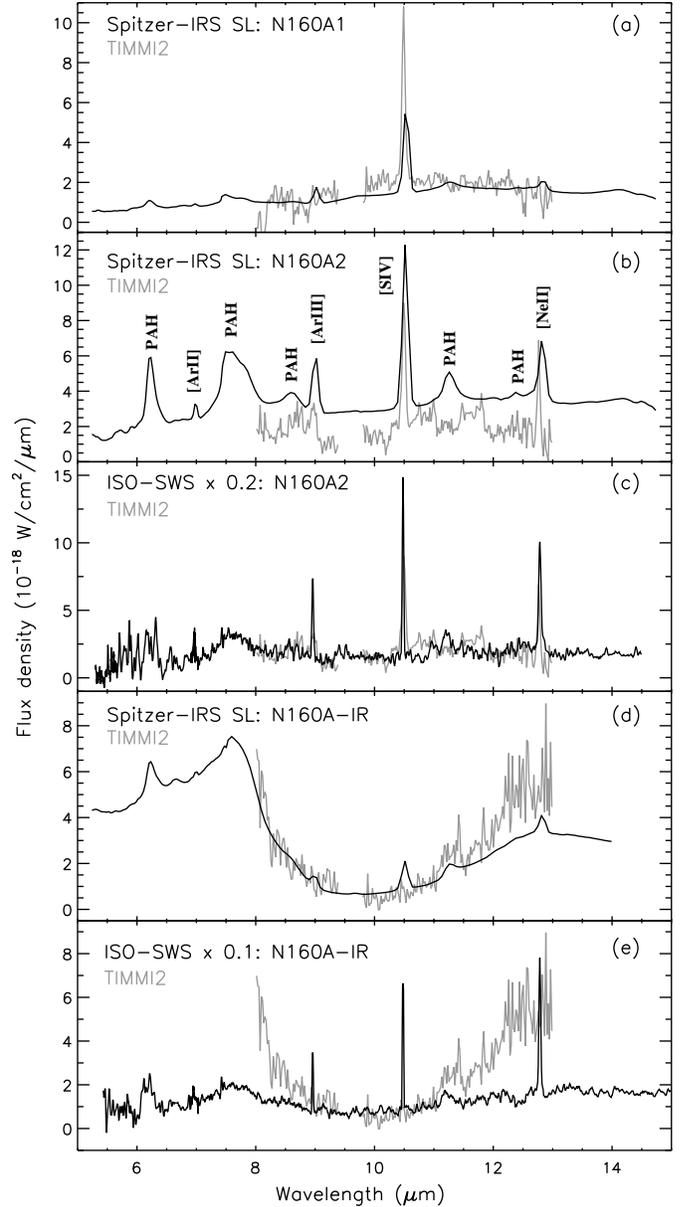}}
    \caption{MIR spectra of N160\,A1, A2 and the YSO N160\,A-IR. Panels {\it a}, {\it b} and {\it d} show the  \spitzer-IRS SL spectra in black and the TIMMI2 spectra in grey. These \spitzer-IRS\ spectra are not scaled down.  Panels {\it c} and {\it e} compare the 
      TIMMI2 spectra of N160\,A2 and N160A-IR (in grey) with the (scaled down) ISO-SWS AOT~01
      spectra centered at these positions (in black).}      
         \label{fig:n160:sp}
   \end{figure}

In terms of molecular features, the MIR spectra of many \HII\ regions are dominated by the well known PAH emission features at 3.3, 6.0, 6.2, 7.7, 8.6, 11.0, 11.2 and 12.7\,\um. Two silicate bands centered around 9.7\,\um\ and 18\,\um\ can also be present.

The TIMMI2, \spitzer-IRS and ISO-SWS line fluxes, and the upper limits of the non-detected
lines, are listed in Table~\ref{tbl:fluxes}. The PAH band fluxes are reported in Table~\ref{tbl:pah}.

\subsubsection{N88\,A}
\label{sect:n88:sp}

Figure~\ref{fig:n88:sp} shows the TIMMI2 and \spitzer-IRS SH spectra of
N88\,A. The spectral appearance of the ISO-SWS spectrum matches that of
the \spitzer-IRS\ spectrum and
therefore, we do not show it.

N88\,A is resolved by TIMMI2 with a FWHM along the spatial direction of
3\farcs0 (see Sect.~\ref{timmi2:obs}), higher than the TIMMI2 slit
width of 1\farcs2. Hence, slit losses are expected. Assuming that the
source is Gaussian and that the slit is perfectly centered on the
source peak, the slit might be registering about 40\% of the total
emission of N88\,A. Assuming this slit loss, the total \SIV\ line flux
associated with N88\,A would be $\sim 85\times10^{-20}$~W\,cm$^{-2}$, in
good agreement with the fluxes measured by \spitzer-IRS\ and ISO-SWS. The same
holds for the dust continuum flux. This indicates that, in spite of
its larger aperture, the fluxes observed by \spitzer-IRS, and even ISO-SWS,
come exclusively from the compact source N88\,A. 
This is expected since N88\,A is an isolated object with no known, adjacent infrared sources. Thus, it can be said that \spitzer\ observations do not spatially resolve N88\,A.

It is interesting to note the lack of PAH emission features in the TIMMI2 spectrum. The \spitzer-IRS spectrum, on the contrary, clearly exhibits a 11.2 \um\ PAH
band. Other PAH bands at e.g.  11.0 and 12.8 \um, and the PAH emission
between 15--20\,\um, are however not present in this \spitzer\ spectrum. In addition, 
hydrogen recombination lines at 11.3 and 12.4 \um\ are detected.

\subsubsection{N160\,A}
\label{sect:n160:sp}


\paragraph{The HEBs A1 and A2}
As it can be seen in the top three panels of Fig.~\ref{fig:n160:sp} (and
in Fig.~\ref{fig:n160:irs}), the spectra of A1 and A2 are the typical
spectra found in  \HII\ regions, characterized, as discussed above, by emission from
fine-structure lines and PAHs on top of a dust continuum.

\begin{table*}
\caption{Line fluxes in units of $10^{-20}$\,W cm$^{-2}$.}
\label{tbl:fluxes}
\begin{center}
\begin{tabular}{l c c c c c l l c} 
\hline\hline\\[-5pt]
Line/ratio & $\lambda$\,(\um) & N88\,A & N160\,A1 & N160\,A2 & N160\,A-IR &
Instrument & Aperture (\arcsec) & Notes\\[2pt]
\hline\\[-5pt]

 \ArII  & 7.0  &      & $0.7\pm0.1$ & $5.2\pm0.5$ & $1.2\pm0.4$ & \spitzer-IRS (SL)& $\alpha\times\alpha~^{\rm (c)}$ \\
        &      & $<5$ & $6\pm2$     & $<6$        & $<6$        & ISO-SWS          & $14\times20$   & 1, 2\\[2pt]

\hline\\[-5pt]
 \ArIII & 9.0  & $<10$     & $<17$       & $<22$      & $<16$       & TIMMI2           & $1.2\times\beta~^{\rm (a)}$ & \\ 
        &      &           & $8.1\pm0.5$ & $32\pm5$   & $4.1\pm0.2$ & \spitzer-IRS (SL)& $\alpha\times\alpha~^{\rm (c)}$ & \\
        &      & $10\pm5$  & $70\pm10$   & $100\pm15$ & $64\pm10$   & ISO-SWS          & $14\times20$  & 1, 2\\[2pt]

\hline\\[-5pt]
 \SIV   & 10.5 & $34\pm2$  & $57\pm3$   & $46\pm4$   & $<12$        & TIMMI2            & $1.2\times\beta~^{\rm (a)}$ & \\
        &      &           & $56\pm4$   & $111\pm3$   & $14.8\pm0.4$ & \spitzer-IRS (SL)& $\alpha\times\alpha~^{\rm (c)}$ & \\
        &      & $66\pm21$  &           & $122\pm61$ &              & \spitzer-IRS (SH) & $4.7\times11.3$  & \\
        &      & $79\pm12$ & $290\pm45$ & $205\pm30$ & $170\pm25$   & ISO-SWS           & $14\times20$  & 1, 2\\[2pt]

\hline\\[-5pt]
 \NeII  & 12.8 & $<16$     & $<16$       & $28\pm7$   & $<36$        & TIMMI2           & $1.2\times\beta~^{\rm (a)}$ & \\
        &      &           & $7.0\pm0.8$ & $45\pm3$   & $11.4\pm0.8$ & \spitzer-IRS (SL)& $\alpha\times\alpha~^{\rm (c)}$ & \\
        &      &$3.3\pm0.5$&             & $59\pm10$  &              & \spitzer-IRS (SH)& $4.7\times11.3$  & \\
        &      & $<5$      & $112\pm35$  & $200\pm40$ & $245\pm45$   & ISO-SWS          & $14\times27$   & 1, 2\\[2pt]

\hline\\[-5pt]
 \NeIII & 15.5 & $71\pm3$  &            & $210\pm37^{\rm (b)}$    & & \spitzer-IRS (SH) & $4.7\times11.3$ & \\
        &      &           & $160\pm70$ & & $96\pm8$                & \spitzer-IRS (LL) & $10.2\times10.2$ & \\
        &      & $90\pm21$ & $440\pm70$ & $370\pm65$ & $575\pm100$& ISO-SWS           & $14\times27$  & 1, 2\\[2pt]

\hline\\[-5pt]
 \SIII  & 18.7 & $22\pm1$  &            & $154\pm22$ &            & \spitzer-IRS (SH)& $4.7\times11.3$ & \\ 
        &      &           & $110\pm20$ & & $68\pm5$              & \spitzer-IRS (LL) & $10.2\times10.2$ & \\
        &      & $<11$      & $320\pm50$ & $270\pm50$ & $215\pm40$ & ISO-SWS          & $14\times27$  & 1, 2\\[2pt]

\hline\\[-5pt]
 \SIII  & 33.5 &          & $98\pm14$     &            & &\spitzer-IRS (LL) & $10.2\times10.2$ & \\
        &      &$12\pm3$  &  & $154\pm22$ &            & \spitzer-IRS (LH)& $11.1\times22.3$       & \\
        &      & $<25$      & $640\pm320$& $600\pm160$& $545\pm140$& ISO-SWS          & $20\times33$  & 1,2\\[2pt]

\hline\\[-5pt]
 \NeIII & 36.0 & $8\pm3$   &            & $37\pm12$  &            & \spitzer-IRS (LH)& $11.1\times22.3$ & \\
        &      & $<21$    & $70\pm35$  & $73\pm20$  & $72\pm20$  & ISO-SWS          & $20\times33$  & 1, 2\\[2pt]

\hline\\[-5pt]
\NeIII/\NeII & 15.5/12.8 & $21\pm3$ &            & $4\pm1$     &   & \spitzer-IRS (SH) & $4.7\times11.3$\\
             &           & $>18$    & $4\pm1$    & $1.8\pm0.5$ & $2.3\pm0.6$ & ISO-SWS & $14\times27$\\ [2pt]

\hline\\[-5pt]
\SIV/\NeII & 10.5/12.8 &  $>2$    & $>4$        & $1.6\pm0.4$ &             & TIMMI2 &  \\
           &           &          & $8\pm1$     &             & $1.3\pm0.1$ & \spitzer-IRS (SL)\\
           &           & $20\pm7$ &             & $2\pm1$     &             & \spitzer-IRS (SH)\\
           &           & $>16$    & $2.6\pm0.9$ & $1.0\pm0.2$ & $0.7\pm0.2$ & ISO-SWS \\[2pt]

\hline
\end{tabular}
\end{center}
$^{\rm (a)}$ 
$\beta$(N88\,A)=4\farcs5,
$\beta$(N160\,A1)=4\farcs7,
$\beta$(N160\,A2)=8\farcs0 and   
$\beta$(N160\,A-IR)=3\farcs5.
$^{\rm (b)}$ This line flux might be uncertain, see Sect.~\ref{spitzer:obs}.
$^{\rm (c)}$~$\alpha$(N160\,A1)=3\farcs7,
$\alpha$(N160\,A2)=7\farcs4 and   
$\alpha$(N160\,A-IR)=5\farcs5.
Notes--(1) The AOT\,02 ISO-SWS line fluxes of N160\,A1 have been published by \cite{vermeij:data}. For N88\,A and N160\,A2, we use instead the 
line fluxes we have directly measured from the AOT\,01 ISO-SWS spectra centered on these sources.
(2) The full ISO-SWS spectrum centered at the position of N160\,A-IR has been published by \cite{henning98}. 
\end{table*}

Both sources are resolved by TIMMI2, with a FWHM along the spatial
direction (see Sect.~\ref{timmi2:obs}) higher than the TIMMI2 slit
width. Again, slit losses are expected, in particular for A2 with a
FWHM of 5\farcs3. In principle, one could expect that the TIMMI2 slit
is registering about 40\% of the total emission in the case of A1, and
only 25\% in the case of A2, assuming that the sources are Gaussian. However, the TIMMI2 and the
\spitzer-IRS spectra of these two sources match very well (see panels {\it a} and {\it b} of Fig.~\ref{fig:n160:sp}) suggesting
that slit losses are not very significant, at least not for the dust
continuum emission.  A comparison with the scaled down ISO-SWS spectrum
of N160\,A2 is shown in the middel panel of Fig.~\ref{fig:n160:sp}.
The large difference 
in the flux level of the continuum between the TIMMI2 and the ISO-SWS spectra
reflects the effect of the much larger ISO-SWS aperture, and
the extensiveness and complexity of the source and its surroundings.

 
Aperture differences are also evident in the line fluxes measured by each instrument (see Table~\ref{tbl:fluxes}).
In the case of A1, for instance, the
\SIV\ line fluxes measured by TIMMI2 and \spitzer\
agree with each other, but they only account for 20\% of the \SIV\ line
flux measured by ISO. 
For A2, the \SIV\ line flux measured by
\spitzer\ is in between those obtained with TIMMI2 and ISO,
which clearly reflects the effect of the extended emission associated
to this whole area. In particular, the \SIV\ line flux measured by TIMMI2  accounts for approximately
40\% of the \SIV\ line flux measured by \spitzer. This percentage is higher than the expected 25\% (see above) and hence, we are confident that the smaller aperture of \spitzer-IRS  (i.e. 4\farcs7$\times$11\arcsec)
is indeed tracing most of the ionized gas associated with A2. 

When looking at the \NeIII/\NeII\ 15.2/12.8 \um\ line ratios measured for N160\,A (see again
Table~\ref{tbl:fluxes}), aperture effects are also evident. Note, for instance, how this ratio (and also the ground-based \SIV/\NeII\ 10.5/12.8\,\um\ ratio) changes depending on the position and/or aperture. Basically, these ratios tend to decrease with aperture size. This has a clear impact if one
wants to use this ratio as a diagnostic for the effective temperature
of the central ionizing star or stars.  The \NeIII/\NeII\ line ratio
is highly sensitive to the effective temperature of the ionizing star \citep[see
e.g.][]{mokiem04,morisset04}, and an incorrect line ratio could lead to
a wrong interpretation in terms of the stellar content.
When sampling a single \HII\ region, the use of pencil beam spectra might give a --sometimes-- wrong impression of the ionization characteristics of the central star. Essentially, the low ionization stage of an ion is undersampled for lines of sight towards the center of the region, and the other way around for lines of sight away from the center. In this sense, it is much better to use observations that encompass the whole \HII\ region. However, N160\,A is not a single \HII\ region. It harbors two HEBs (each containing at least one massive O star) and the main ionizing stars (from which the hottest could be an O6.5\,V) appear to be part of a central cluster which extends over several arcseconds. This cluster is located between A1 and A2 (see the left panel of Fig.~\ref{fig:rgb}). The ionization structure of N160\,A is, therefore, highly complex. The variations we see in the \NeIII/\NeII\ (and \SIV/\NeII) line ratio, with the highest value found close to the HEBs, suggest that the two HEBs are ionized by very hot stars, while the extended emission is ionized by cooler stars that are likely located in the central cluster.  These two HEBs could be ionized, in fact, by hot O4--O5\,V stars, as estimated from their 4.8\,GHz radio continuum emission \citep{martin:atca:mc}.

Finally, N160\,A1 exhibits PAH emission bands at 6.0, 6.2,
6.8, 7.7, 8.6, 11.2 and 12.7 \um\ (see panel {\it a} of Fig.~\ref{fig:n160:sp}). 
It has a remarkably strong 6.0~\um\,
band which starts already near 5.83 \um.  The 6.2/11.2 and 7.7/11.2
band ratios (see Table~\ref{tbl:pah}) are slightly lower than
those of N160\,A2, but similar to those found towards LMC \HII\
regions \citep{vermeij02}. In addition, the band profiles of the main
PAH bands belong to class A and are, therefore, consistent
with the typical PAH characteristics found in \HII\ regions
\citep{peeters02}. The
top of the 7.7 \um\ complex, however, is fairly sharp and emission seems to
be lacking from the 7.8 \um\, component. This could possibly be due to
the less reliable fluxes at the order edges and to mismatches between the different orders of \spitzer-IRS SL (one order overlap falls exactly at
the position of the 7.7 \um\ complex). 

N160\,A2, on the other hand, exhibits PAH emission bands at 5.75, 6.0, 6.2, 7.7, 8.6,
11.0, 11.2 and 12.7 \um\ (see panel {\it b} of Fig.~\ref{fig:n160:sp}).
The profiles of the main
PAH bands (at 6.2, 7.7, 8.6 and 11.2 \um)  belong to class A.
As for the fine-structure lines, aperture
effects are apparent. The PAH band fluxes measured by
\spitzer\ are considerably smaller than the ISO ones. In particular, they
only account for about 24\% and 15\% of the, respectively, 7.7
and 11.2 \um\, PAH fluxes measured by ISO. This percentage is roughly similar to the
corresponding aperture ratio, suggesting that the PAH emission is very homogeneous. 
In addition, there exists a
significant difference in the 7.7/(11.0+11.2) band ratio
between the \spitzer\ and the ISO data. Given that the 7.7 and 11.2 \um\
emission is dominated by PAH cations and neutrals respectively, this
suggests a higher neutral to cation ratio in the larger ISO-SWS
aperture.  
In general, the  7.7/(11.0+11.2) band ratio is similar to those found in LMC
\HII \ regions by \citet{vermeij02}, and is on the lower half of the
range observed towards Galactic \HII \ regions \citep{hony01}. The 12.7/11.2 band
ratio, on the contrary, is more similar to the ratios observed towards Galactic intermediate
mass star-forming regions and Planetary Nebulae than towards Galactic \HII \
regions. 
This suggests the presence of more compact,
smooth-edged PAHs towards N160\,A2. However, a much more detailed
study of the CH out-of-plane bending modes in Magellanic Cloud \HII\ regions is
warranted.

    \begin{figure}[!h]
   \centering \resizebox{\hsize}{!}{
   \includegraphics{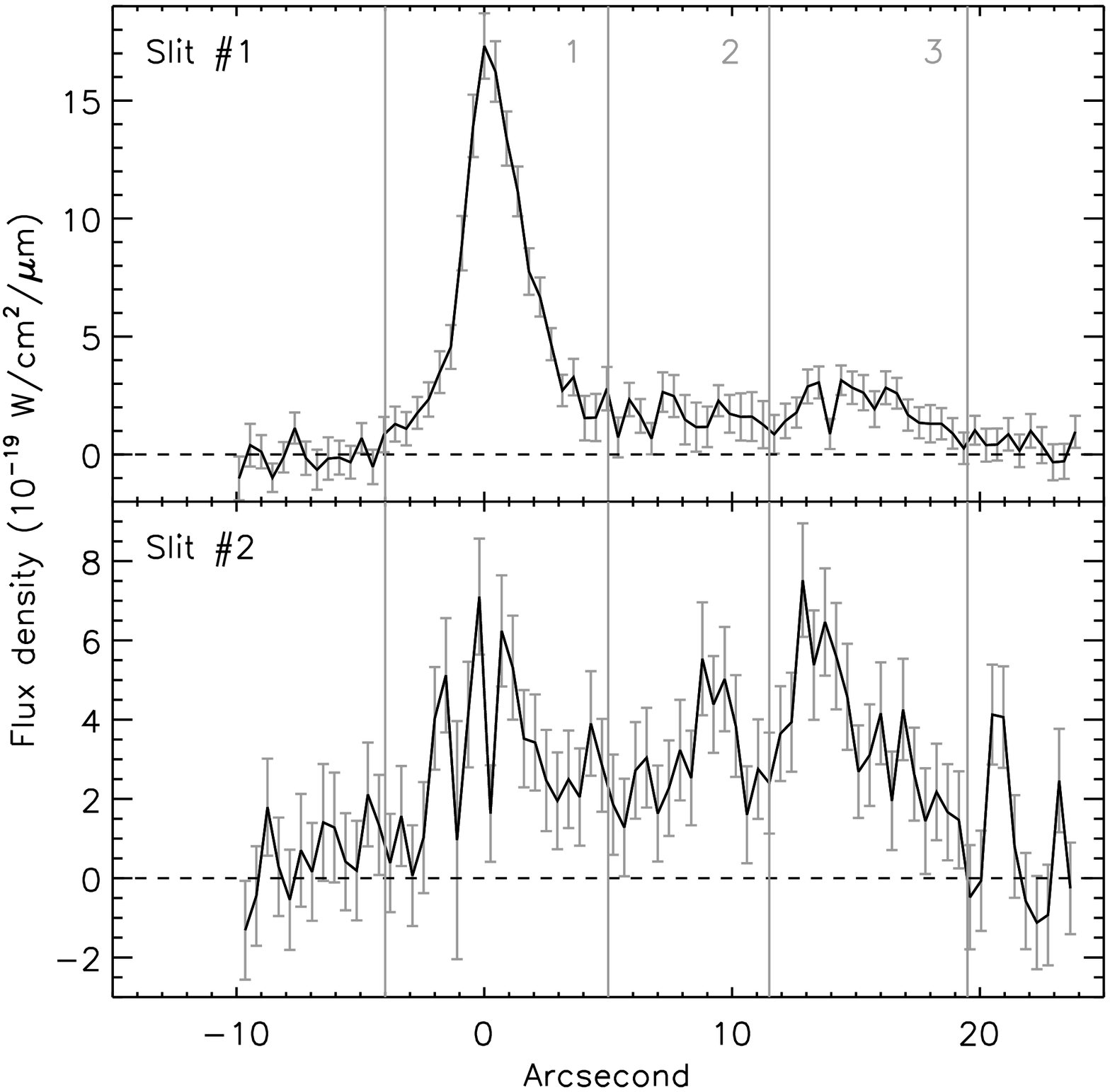}}
      \caption{Spatial variation  of the \SIV\ 10.5\,$\mu$m line peak across slit \#1 (s1) and 
       slit \#2 (s2) of TIMMI2. North is to the right. 
       Three distinct zones (1, 2 and 3) are indicated (see also the left panel of Fig.~\ref{fig:rgb}).}
         \label{fig:n160a:siv}
   \end{figure}
  
  \begin{figure}[!ht]
   \centering \resizebox{\hsize}{!}{
   \includegraphics{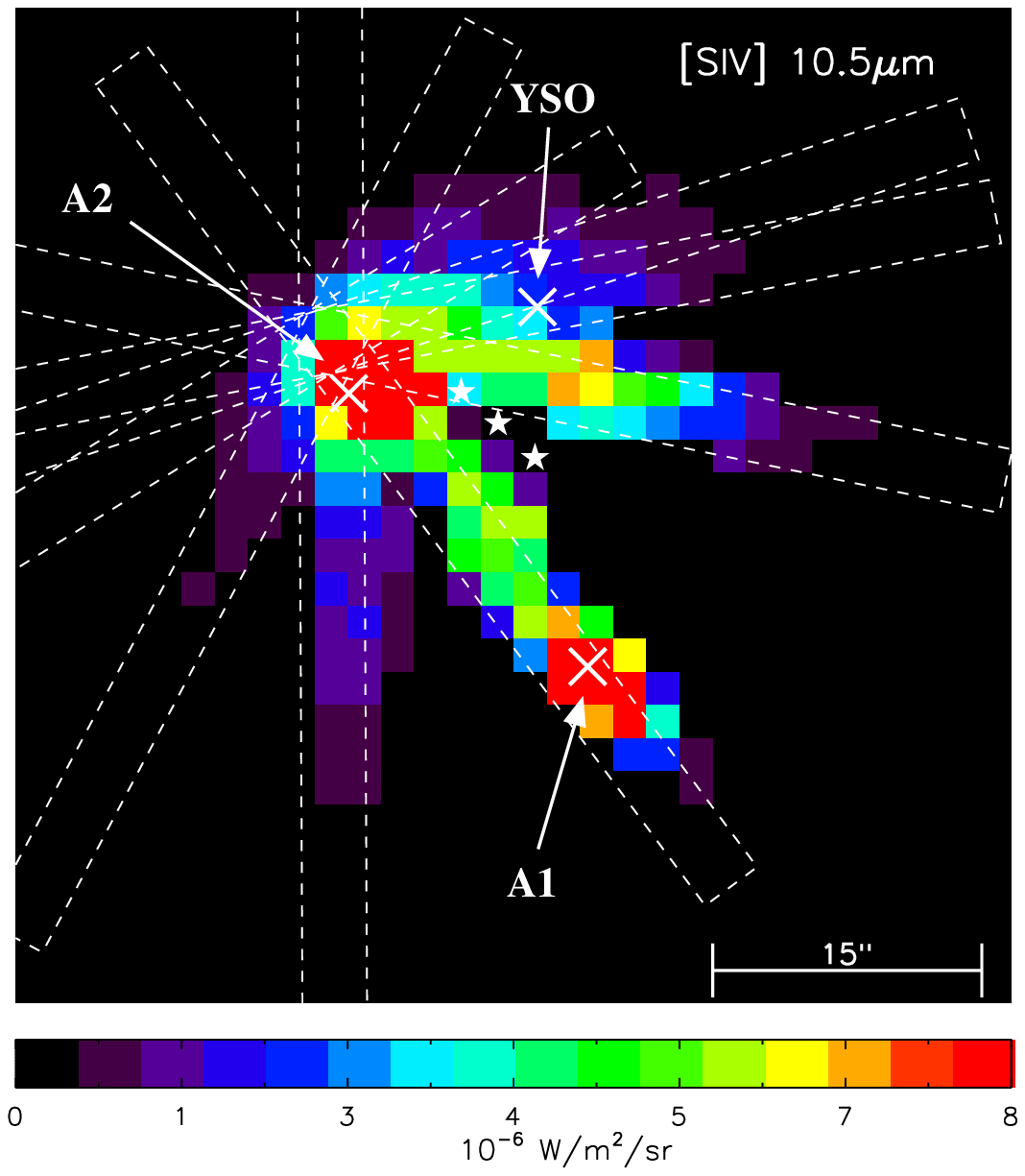}}
      \caption{\SIV\ 10.5 \um\ line flux map of N160\,A. Its patchy appearance is the result of the limited spatial coverage. The dashed lines are a few examples of the \spitzer-IRS SL apertures. The brightest stars of the central cluster \citep[stars \#24, \#35 and \#58 of][]{heydari02b} are also  indicated [see electronic edition of the Journal for a color version of this figure].}
         \label{fig:n160a:map}
   \end{figure}

\paragraph{The extended emission}
TIMMI2 slits s1 and s2 registered faint and extended \SIV\ emission.  Extended emission from other atomic lines was not detected with TIMMI2. Figure~\ref{fig:n160a:siv} shows the variation of the \SIV\ line peak flux across these slits from south to north. This emission practically extends across the 20\arcsec\ of separation between the YSO and A1, and is most likely associated with the central star cluster located between A1 and A2. This cluster extends over an area $\sim9\arcsec\times8\arcsec$ and its brightest component, if it were a main sequence star, could be, as mentioned in Sect.~\ref{desc:n160a}, an O6.5\,V.

A more complete map of this extended \SIV\ line emission is shown in Fig.~\ref{fig:n160a:map}. This map has been created using the multiple \spitzer-IRS SL observations that are available. It is clear that the \SIV\ emission is stronger at the positions of the A1 and A2 components, while it is practically null at the position of the YSO. It is also evident that there exists an extended component towards the west and south-west.

\paragraph{The YSO N160\,A-IR}
The spectrum of N160\,A-IR, shown in the lower two panels of Fig.~\ref{fig:n160:sp}, is dominated by a deep silicate absorption, indicative of the high extinction (around 70~mag, see Sect.~\ref{dusty}) existent towards this object. The TIMMI2 spectrum lacks emission from fine-structure lines.
Once again, the overall agreement between the TIMMI2 and \spitzer\ observations is striking, in particular in the case of the continuum emission/absorption. On the other hand, it is clear that the dominant contribution to the ISO-SWS aperture is the ionized gas surrounding the YSO. 
Even for the smaller \spitzer-IRS SL aperture (5\farcs5$\times$5\farcs5), a contribution of the surrounding ionized gas is present in the form of fine-structure lines. In the case of the TIMMI2 observations, one might expect slit losses of about 50\%. However, the TIMMI2 spectrum agrees remarkably well with the photometry of \cite{epchtein84} obtained through a 10\arcsec\ aperture (see Fig.~\ref{fig:sed}).

\begin{table}[!tp]
\caption[PAH fluxes]{PAH fluxes in units of 10$^{-20}$\,W\,cm$^{-2}$.}
\label{tbl:pah}
  \begin{center}
    \leavevmode
\begin{tabular}{l@{\hspace{2pt}}cccccl}
\hline\hline\\[-5pt]
Source & 6.0+6.2 & 7.7 & 8.6 & 11.0+11.2 & 12.7 & Instrument\\ 
\hline \\[-5.pt]
N88\,A   &    --   &  --  &  -- & 0+15$^{\rm (a)}$ &  -- & IRS-SH\\
N160\,A1 &    11   & 22   & 1.4 & 12       &  {\it y} & IRS-SL\\
N160\,A2 &    77   & 165  & 18  & 45       &  -- & IRS-SL\\
         &   --    &  --  &  -- & 5+39     & 14  & IRS-SH\\
         &     {\it y}   &  693 & {\it y}   & 60+233   & {\it y}   & ISO-SWS\\
\hline \\[-0.5cm]
\end{tabular}
  \end{center}

{\it y} means that the band is present but it is very weak.
$^{\rm (a)}$ No 11.0 \um\, band present.
Notes--Integration limits are [7.07,8.2] and [8.2, 8.83] for the 7.7
and 8.6 \um\ PAH bands, respectively. The integration limits for the
11.0+11.2 \um\ bands are [10.85, 11.6].

\end{table}

\begin{table}[!tp]
\caption[Abundances]{Elemental abundances.}
\label{tbl:ab}
  \begin{center}
    \leavevmode
\begin{tabular}{lccc}
\hline\hline\\[-5pt]
& N\,88A & N160\,A1& N160\,A2\\ 
\hline \\[-5.pt]
Ne/H & $1.7\pm0.1$ & $7\pm3$ & $4.9\pm0.8$\\
($\times 10^{-5}$) & $0.94\pm0.04$\,$^{\rm (a)}$ & $6.7\pm0.9$\,$^{\rm (a)}$ & $5.73\pm0.02$\,$^{\rm (a)}$\\
& 2.1\,$^{\rm (c)}$ & 6.33\,$^{\rm (b)}$ & 4.48\,$^{\rm (b)}$\\
& 2.2\,$^{\rm (d)}$\\[4pt]

S/H & $1.2\pm0.2$ & $7\pm1$ & $3.6\pm0.7$\\
($\times 10^{-6}$)& $1.2\pm0.3$\,$^{\rm (a)}$ & $7.8\pm0.5$\,$^{\rm (a)}$ & $6.5\pm0.4$\,$^{\rm (a)}$\\
& 3.2\,$^{\rm (c)}$ & $7\pm2$\,$^{\rm (b)}$ & $5\pm1$\,$^{\rm (b)}$\\
& 2.2\,$^{\rm (d)}$\\[4pt]

Ar/H & --          & $5.2\pm0.3$ & $8\pm1$\\
($\times 10^{-7}$) & & $14.8\pm0.4$\,$^{\rm (a)}$ & $15.1\pm0.2$\,$^{\rm (a)}$\\
\hline \\[-0.5cm]
\end{tabular}
  \end{center}
References: $^{\rm (a)}$ infrared: \cite{vermeij:analysis};
$^{\rm (b)}$ optical: \cite{heydari86};
$^{\rm (c)}$ optical: \cite{kurt99};
$^{\rm (d)}$ optical: \cite{testor03}.
\end{table}

In addition to the silicate absorption bands at 9.7 and 18\,\um, the \spitzer\ spectrum shows PAH emission at 6.2, 7.7 and 11.2\,\um, and even marginally at 8.6 \um\ (see panel {\it d} of Fig.~\ref{fig:n160:sp}). 

\subsection{Physical conditions and abundance calculations}

The electron density can be determined from the ratio of two atomic fine-structure lines of the same ionic species when they are emitted from levels with nearly the same excitation energy \citep[e.g.][]{rubin94}. This ratio is sensitive to gas densities approximately in between the critical densities of each line. High density gas can be probed using line ratios such as \SIII\ 33.5/18.7 and \NeIII\ 36.0/15.5~\um. On the other hand, ionic abundances can be determined from the measured strengths of the fine-structure and \HI\ recombination lines \citep[e.g.][]{rubin88}.

We will now determine the physical conditions of the ionized gas (in terms of electron density and elemental abundances) in N88\,A, N160\,A1 and N160\,A2. The final abundances of Ne, S and Ar are listed in Table~\ref{tbl:ab}, where they are compared with previous optical and infrared determinations.

\subsubsection{N88\,A}
\label{ab:n88a}

We use directly the line fluxes measured by \spitzer-IRS. De-reddening is not an issue because of the low extinction at these wavelengths. Note that the extinction in e.g. the near-infrared is only 2\% of the attenuation suffered in the optical (which is around 1.5) when applying the SMC extinction law obtained by \cite{gordon03}. 

Assuming an electron temperature of 14\,000\,K for N88\,A \citep[e.g.][]{kurt99,vermeij:analysis,testor03}, the observed \SIII\ 33.5/18.7 line ratio gives an electron density of $2275 (+390,-310)$ cm$^{-3}$. This density is comparable to the one obtained by \cite{vermeij:analysis} using ISO-SWS (2200\,cm$^{-3}$), and to the rms electron density (3500\,cm$^{-3}$) derived from radio continuum observations (\citeauthor{indebetouw04} \citeyear{indebetouw04} measured a flux density of 91 mJy at 6 cm) using Eq. A.2.5 of \cite{panagia78} and a source diameter of 3\arcsec\ (see Sect.~\ref{timmi2:obs}). The observed \NeIII\ 36.0/15.5 ratio, on the other hand, indicates a density $\lesssim 10^4$\,cm$^{-3}$.

In the calculation of ionic abundances, we use the H$\beta$ line flux ($2.6\times10^{-11}$~erg\,s$^{-1}$\,cm$^{-2}$) predicted directly from radio continuum observations. It is estimated using Eq. 10 of \cite{martin:atca:mc}. When this H$\beta$ line flux is compared with the line flux measured by \cite{heydari99c}, we get a visible extinction $A_{\rm V}\sim 1.8$ which is in agreement with the values obtained by previous authors (see Sect.~\ref{desc:n88a}). We also assume an electron density of 3500\,cm$^{-3}$. This density is lower than the critical densities of the lines involved in the calculations and hence, is not a decisive factor.
We obtain Ne$^+$/H$^+ = 1.5(0.2)\times10^{-6}$,
Ne$^{++}$/H$^+ = 1.5(0.1)\times10^{-5}$,
S$^{++}$/H$^+ = 7.2(0.3)\times10^{-7}$ and
S$^{+3}$/H$^+ = 5(2)\times10^{-7}$, where the uncertainties are given within brackets.

The total Ne abundance can be determined by adding the contributions of Ne$^+$ and Ne$^{++}$.
The final Ne/H we estimate agrees well with the optical measurements. It is, however, higher than the abundance obtained by \cite{vermeij:analysis}, but these authors only give the contribution of Ne$^{++}$ since they could not determine the ionic abundance of Ne$^{+}$.

In the case of S, adding the contributions of S$^{++}$ and S$^{3+}$ only accounts for $\sim$50\% of the expected sulphur abundance, i.e. about 0.2 times the solar S abundance \citep[where we use the solar S abundance given by][]{asplund05}. \cite{testor03} estimate a S$^{++}$/H$^+$ abundance only 30\% higher than the value we calculate and an insignificantly low S$^{+}$/H$^+$. It might then be inferred that 50\% of the S abundance must be in the form of S$^{+4}$, which needs energies around 47\,eV in order to be produced. In fact, the optical estimates, which use ionization correction factors to account for unobserved ions, give higher S abundances.

\subsubsection{N160\,A1}
\label{ab:n160a1}

We use directly the line fluxes measured by \spitzer-IRS\ (SL and LL modes). For an electron temperature of 10\,000~K \citep{heydari86, vermeij:analysis}, the observed \SIII\ 33.5/18.7 line ratio gives an electron density of $780 (+410,-250)$~cm$^{-3}$, similar to the density estimated by \cite{vermeij:analysis} using ISO-SWS (around 450\,cm$^{-3}$), but somewhat lower than the rms electron density of 2550\,cm$^{-3}$. This rms electron density is derived from radio continuum observations assuming a source diameter of 3\farcs1 (see Sect.~\ref{timmi2:obs}) and using the 6~cm flux density measured by \cite{indebetouw04}.

In the calculation of ionic abundances, we use the H$\beta$ line flux predicted directly from radio continuum observations, and assume an electron density of 2550\,cm$^{-3}$ --this density is again lower than the critical densities of the lines involved in the calculations. This H$\beta$ line flux ($1.7\times10^{-11}$~erg\,s$^{-1}$\,cm$^{-2}$) is similar to the de-reddened H$\beta$ line flux estimated by \cite{heydari02b}.
For Ne we obtain
Ne$^{+}$/H$^+ = 6.1(0.7)\times10^{-6}$ and
Ne$^{++}$/H$^+ = 6(3)\times10^{-5}$. For S,
S$^{++}$/H$^+ = 6(1)\times10^{-6}$ and
S$^{+3}$/H$^+ = 7.8(0.6)\times10^{-7}$. And finally, for Ar,
Ar$^{+}$/H$^+ = 3.5(0.5)\times10^{-8}$ and
Ar$^{++}$/H$^+ = 4.9(0.3)\times10^{-7}$. 

The abundances of Ne, S and Ar are given in Table~\ref{tbl:ab}. The abundances of Ne and S agree well with the optical and infrared estimates. The Ar abundance we obtain, however, is about 3 times lower than the one estimated by \cite{vermeij:analysis} using ISO-SWS.
ISO presented a clear advantage over other instruments when estimating elemental abundances: fine-structure lines and strong \HI\ emission lines (e.g. Br$\alpha$ and Br$\beta$ at 4.05 and 2.62\,\um, respectively) were observed using roughly the same aperture. Since in the estimate of ionic abundances we are assuming that the volume occupied by both ionic gases  (i.e. H$^+$ and X$^{+i}$, where X is a given element) is the same, aperture mismatches clearly have an effect on the results. 
This effect is more pronounced in the case of Ar$^+$, probably because this ion has a lower ionization potential (27.6 eV) than others such as S$^{++}$ (34.8~eV) and Ne$^{+}$ (41.0~eV). The volume occupied by Ar$^+$ is thus largely extended, and we are likely underestimating the \ArII\ 7.0\,\um\ line flux.

\subsubsection{N160\,A2}
\label{ab:n160a2}

As it has been done previously, we use directly the line fluxes measured by \spitzer-IRS. In particular, we use those measured in the SH mode, which cover a wider wavelength range and, when there is an overlap, agree very well with the line fluxes measured in the SL mode. 
Assuming an electron temperature of 10\,000\,K \citep[e.g.][]{heydari86}, the observed \SIII\ 33.5/18.7 line ratio gives a much lower electron density ($< 200$ cm$^{-3}$) than the rms electron density derived from the 6~cm continuum observations, i.e. 1850~cm$^{-3}$, estimated by using the 130 mJy measured by \cite{indebetouw04} and a source diameter of 5\farcs3 (see Sect.~\ref{timmi2:obs}). We believe this is because of the different apertures through which both \SIII\ lines are observed (see Table~\ref{tbl:fluxes}); we might perfectly be overestimating the \SIII\ 33.5~\um\ line flux associated with A2. The observed \NeIII\ 36.0/15.5 indicates a density $\lesssim 10^4$\,cm$^{-3}$.

In the calculation of ionic abundances, we use the H$\beta$ line flux ($4.6\times10^{-11}$~erg\,s$^{-1}$\,cm$^{-2}$) predicted directly from radio continuum observations. 
When compared with the H$\beta$ line flux measured by \cite{heydari02b}, we get a visible extinction around 2.5, slightly higher than the mean estimation given by these authors ($\sim 1.5$, see Sect.~\ref{desc:n160a}).
We also assume an electron density of 1850\,cm$^{-3}$. This density is again lower than the critical densities of the lines involved in the calculations. 
We obtain Ne$^+$/H$^+ = 1.9(0.3)\times10^{-5}$,
Ne$^{++}$/H$^+ = 3.0(0.5)\times10^{-5}$,
S$^{++}$/H$^+ = 3.0(0.4)\times10^{-6}$ and
S$^{+3}$/H$^+ = 6(3)\times10^{-7}$.
We can as well estimate the Ar$^+$ and Ar$^{++}$ ionic abundances using the \ArII\ and \ArIII\ line fluxes measured by \spitzer-IRS in the SL mode. We obtain
Ar$^{+}$/H$^+ = 1.0(0.1)\times10^{-7}$ and
Ar$^{++}$/H$^+ = 7(2)\times10^{-7}$.

The abundances of Ne, S and Ar are given in Table~\ref{tbl:ab}. The Ne abundance agrees well with the optical and infrared estimates. 
In the case of S, adding the contributions of S$^{++}$ and S$^{3+}$ only accounts for $\sim$50\% of the expected sulphur abundance, similarly to what we find for N88\,A. \cite{heydari86} estimate a S$^{++}$/H$^+$ abundance only 15\% higher than the value we calculate, and an extremely low S$^{+}$/H$^+$ abundance (about $3\times10^{-7}$). On the other hand, for this region the S$^{+3}$/H$^+$ is sufficiently low that one does not expect a significant contribution in the form of S$^{+4}$. Adding the contribution of S$^{+}$, the total S/H abundance we obtain is
$\sim 4\times10^{-6}$, about 0.3 times the sulphur solar abundance \citep{asplund05}, instead of the canonical 0.5 value. This abundance is in between the lower and upper limits determined by \cite{heydari86}.
It is, however, slightly lower than the S abundance obtained by \cite{vermeij:analysis}.

A larger discrepancy is found in the case of Ar. The Ar abundance we obtain is about 2 times lower than the one estimated by \cite{vermeij:analysis} using ISO-SWS. As mentioned previously for N160\,A1 in Sect.~\ref{ab:n160a1}, aperture mismatches are probably the cause of this difference.

\begin{table}[!tp]
\caption{Photometry of N160\,A-IR.}
\label{tbl:sed}
  \begin{center}
    \leavevmode
\begin{tabular}{ccccc}
\hline\hline\\[-5pt]
\multicolumn{1}{c}{$\lambda$} &
\multicolumn{1}{c}{$\Delta\lambda$} &
\multicolumn{1}{c}{Flux density} &
\multicolumn{1}{c}{Band} &
\multicolumn{1}{c}{Reference}\\
(\um) & (\um) & (mJy) \\[5.pt]
\hline \\[-5.pt]
1.25   & 0.30  & $<$0.08            & $J$           & 1\\ 
1.65   & 0.30  & $<$0.24            & $H$           & 1\\
2.15   & 0.32  & 1.1$\pm$0.2        & $K'$          & 1\\
3.56   & 0.75  & 50$\pm$2           & IRAC-3.6\,\um & 2\\
3.76   & 0.70  & 80$\pm$9           & $L$           & 3\\
4.52   & 1.01  & 190$\pm$3          & IRAC-4.5\,\um & 2\\
4.69   & 0.68  & 267$\pm$25         & $M$           & 3\\
5.73   & 1.42  & 547$\pm$6          & IRAC-5.8\,\um & 2\\
7.91   & 2.93  & 852$\pm$22         & IRAC-8.0\,\um & 2\\
8.38   & 0.96  & 845$\pm$272        & $N1$          & 3\\
9.69   & 1.50  & 437$\pm$241        & $N2$          & 3\\
11.88  & 6.53  & $<$123$\times10^2$ & IRAS-12\,\um  & 4\\
12.85  & 1.38  & 3027$\pm$558       & $N3$          & 3\\
18.06  & 4.12  & 6081$\pm$1680      & $Q$           & 3\\
23.80  & 9.12  & $<$113$\times10^3$ & IRAS-25\,\um  & 4\\
50     & 20    & $<$230$\times10^3$ & KAO           & 5\\
60.85  & 25.89 & $<$623$\times10^3$ & IRAS-60\,\um  & 4\\
71     & 19    & $22900\pm100$      & MIPS-70\,\um  & 2\\
100    & 50    & $<$300$\times10^3$ & KAO           & 5\\
102.44 & 39.55 & $<$791$\times10^3$ & IRAS-60\,\um  & 4\\
1.3d3  &       & 199$\pm$14         & SEST          & 1\\[5.pt]

\hline \\[-0.5cm]
\end{tabular}
  \end{center}
References:
(1) \cite{henning98}; 
(2) this work;
(3) \cite{epchtein84};
(4) IRAS Point Source Catalogue v.2.1;
(5) \cite{werner78}.
\end{table}

  \begin{figure*}
   \centering \resizebox{0.9\hsize}{!}{
   \includegraphics{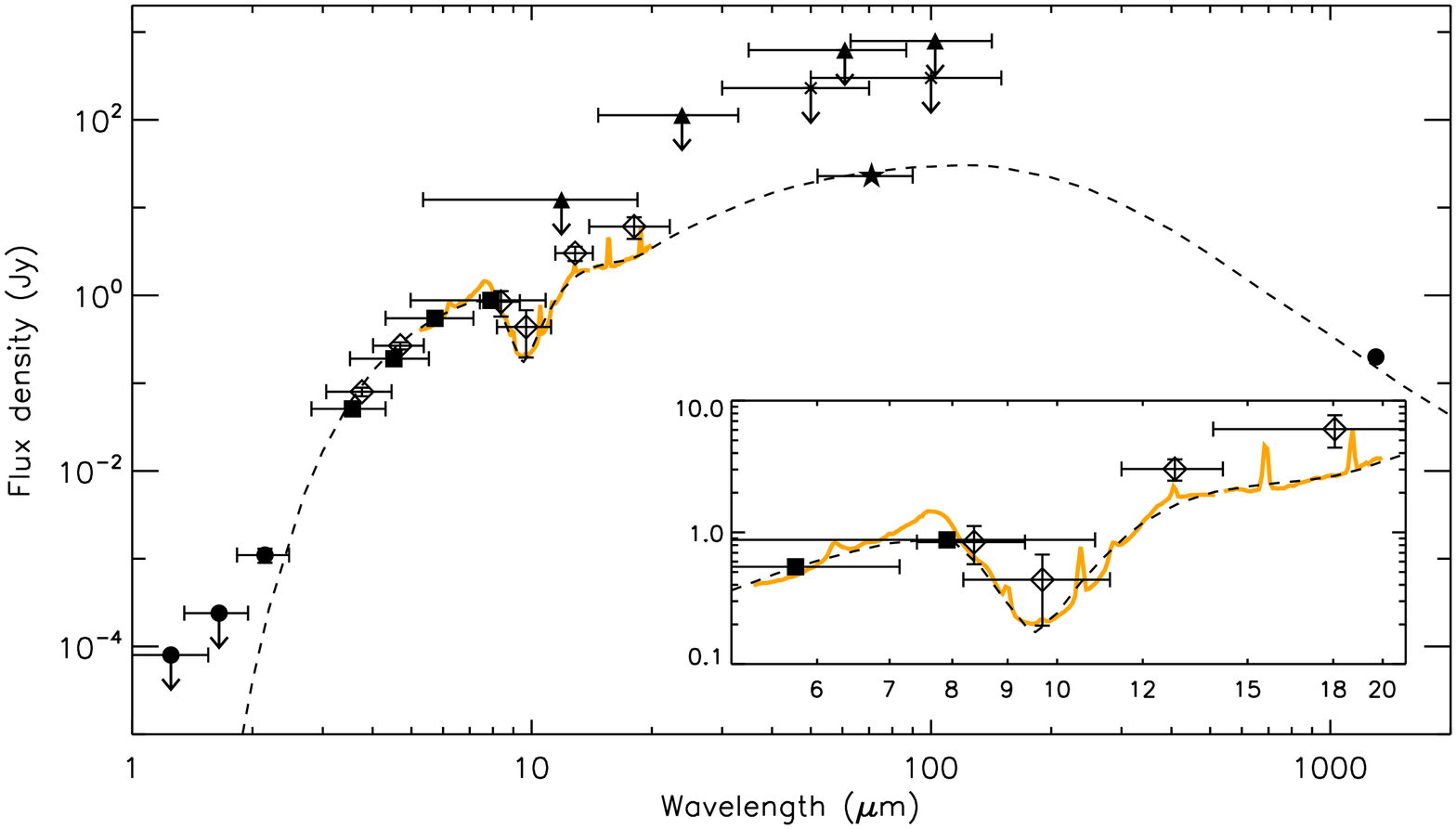}}
      \caption{Spectral energy distribution of N160\,A-IR between 1\,\um\ and 
      1.3\,mm. The \spitzer-IRS SL and LL ($2^{nd}$ order) spectra are plotted in orange.
      The legend of the data is as follows: filled circles are data by 
      \cite{henning98}, open diamonds by \cite{epchtein84} and crosses by 
      \cite{werner78}; filled squares correspond to the \spitzer-IRAC photometry, the filled star to the \spitzer-MIPS photometry, and
      filled triangles are the IRAS fluxes. The best fit obtained with the
      {\sc dusty} code is plotted by a dashed line
      [see electronic edition of the Journal for a color version of this figure].}
         \label{fig:sed}
   \end{figure*}

\subsection{The spectral energy distribution of the massive YSO N160A-IR}
\label{dusty}

Table~\ref{tbl:sed} compiles all the available photometry for N160\,A-IR, where no color corrections are applied. The IRAC photometry has been obtained using an aperture of 7\farcs2. For MIPS, we use a 16\arcsec\ aperture (note that the pixel scale is 4\arcsec\,pixel$^{-1}$) centered on the position of the YSO. The photometry of \cite{epchtein84} was obtained using apertures of 3\farcs3 for the $L$ and $M$ bands, and 10\arcsec\ for the $N1$, $N2$ and $N3$ bands.
Far-infrared KAO flux densities are integrated over 1\arcmin\ and hence are taken as upper limits. We also take IRAS values as upper limits. The beam width of SEST at 1.3\,mm is 22\arcsec. The photometry is plotted, together with the \spitzer-IRS SL and LL (2$^ {nd}$ order) spectra, in Fig.~\ref{fig:sed}. The LL spectrum (2$^ {nd}$ order), which is observed in a 10\arcsec\ aperture, was scaled down by a factor of 1.6 to the level of the SL spectrum, observed in a smaller aperture of 5\farcs5. The 5-20\,\um\ photometry agrees well with the observed spectra. 

The SED of N160\,A-IR has been reproduced with the dust radiative transfer model {\sc dusty} \citep{ivezic99} as an attempt at deriving approximate quantitative information about this YSO. For {\sc dusty} we just have to specify the normalized spectrum of the central star, the dust composition mix and its radial distribution. The code computes the dust temperature radial distribution and the emerging radiation field. The dust composition can be chosen from a variety of grain types, but we decided to use the standard composition \citep{weingartner01}, i.e. silicate and graphite from \cite{draine84}, and amorphous carbon from \cite{hanner88}. The relative proportion of each component is a free parameter. A more detailed spectrum than the one we have is necessary to investigate the presence of other types of grains.  We can, however, assess the presence or absence of silicate, graphite and amorphous carbon from the SED plotted in Fig.~\ref{fig:sed}. The depth of the absorption at 9.7\,\um, for instance, is very sensitive to the relative proportion of silicate. In the absence of silicate, graphite will be responsible for most of the emission below 9\,\um, while amorphous carbon will mostly emit between 20 and 100\,\um\ \citep{plante02}.

We placed a blackbody with an effective temperature of 50\,000~K in the center of a spherically symmetric dust cloud. The result is hardly dependent on the exact value of this temperature within a reasonable range. The geometry of the massive YSO, on the other hand, is almost certainly not spherically symmetric, but the present data are unable to constrain more detailed geometries. We also assume a standard MNR grain size distribution \citep{mathis77}.

The best model fit (dashed line in Fig.~\ref{fig:sed}) uses a mixture with silicates (50\%), graphite (30\%) and amorphous carbon grains (20\%), and a single power law of the form $\rho{\rm (r)} \propto r^{-1.0}$ for the radial profile of the density in the dust cloud. The observations are reproduced with an $A_{\rm V} = 70\pm10$ mag. The dust cloud extends from an inner radius of a few hundred ($\sim200$) AU, at which the dust temperature reaches 590\,K, to an outer radius of about 1 pc (but note that the observable size of the source will depend on the selected wavelength).
The observed SED is well reproduced, specially the silicate feature at 9.7\,\um. This model gives an optical depth of the 9.7\,\um\ silicate of 4.29, and a ratio of the fluxes at 9.7 and 18\,\um\ of $2.7\times10^{-1}$.

 The absolute bolometric luminosity of the central source is the simplest parameter to derive from the model. We obtain a  bolometric luminosity  of $2\times 10^5L_{\sun}$. It is derived by scaling the model SED to fit the 1.3\,mm data point, and using a value for the distance to the LMC of 50 kpc \citep{storm04}. This luminosity suggests a central star of spectral type O6\,V and about $30\,M_{\sun}$ \citep{martins05}.
The fit by \cite{henning98} derived a higher bolometric luminosity ($1.4\times 10^6L_{\sun}$), but their SED was scaled up to the ISO-SWS spectrum centered on the YSO, whose continuum flux level is about 10 times higher than that observed by TIMMI2 and \spitzer-IRS\ (see panel {\it e} of Fig.~\ref{fig:n160:sp}).

\section{Conclusions}
\label{conclusions}

We have presented MIR spectra of two Magellanic Cloud \HII\ regions: N88\,A in the SMC and N160\,A in the LMC. N88\,A is a compact ($\sim3$\arcsec\ in diameter) nebula, but N160\,A is comprised of 3 bright components: two high-excitation blobs, A1 and A2, and a young stellar object, N160\,A-IR. These spectra have been obtained with TIMMI2 on the ESO 3.6 m telescope and the Infrared Spectrograph (IRS) onboard the \spitzer\ Space Telescope.  With these high-spatial resolution data, we have been able to separate for the first time, and from a spectroscopic point of view, the three components of N160\,A. 

The spectra show a wide variety in terms of continuum, lines and molecular bands strength. N88\,A, N160\,A1 and N160\,A2 show a rising dust continuum, PAH bands and strong fine-structure lines. These spectra are typical of \HII\ regions. On the contrary, the YSO present in N160\,A shows a strong dust continuum with deep silicate absorption. 

The influence of spatial resolution is investigated by means of comparing these TIMMI2 and \spitzer-IRS spectra with ISO-SWS observations.  
Observations of N88\,A indicate that the fluxes observed by ISO-SWS and \spitzer-IRS come, in spite of their large apertures, exclusively from the compact 3\arcsec\ source. We derived a \SIII\ electron density of 2275~cm$^{-3}$, and elemental abundances of $\sim1.7\times10^{-5}$ for Ne, and of $\sim1.2\times10^{-6}$ for~S.

Concerning N160\,A1 and N160\,A2, aperture effects are readily observed. In this respect, it is worthwhile to note the large differences observed in, for example, the \NeIII/\NeII\ line ratio, given that this ratio is generally used as a diagnostic for the effective temperature of the central ionizing star(s). This ratio tends to decrease with aperture size, which reflects the complexity of the ionization structure of N160\,A. The variations seen in the \NeIII/\NeII\ line ratio, with the highest value found close to the HEBs, suggest that the two HEBs are ionized by very hot stars, while the extended emission is ionized by cooler stars probably located in the central cluster. This central cluster is situated between A1 and A2, and extends over several arcseconds.

For A1, we derived a \SIII\ electron density of 780 cm$^{-3}$ and elemental abundances of $\sim 7\times10^{-5}$ for Ne, of $\sim 7\times10^{-6}$ for S and of $\sim 5.2\times10^{-7}$ for Ar. For A2, we derived elemental abundances of $\sim 4.9\times10^{-5}$ for Ne, of $\sim 3.6\times10^{-6}$ for S and of $\sim 8.0\times10^{-7}$ for Ar. In the case of A2, the derivation of the electron density using the \SIII\ lines suffers from aperture mismatches. We also observed weak and extended \SIV\ emission with TIMMI2, which is most likely related to the central stellar cluster located between A1 and A2. The PAH emission bands in both A1 and A2 show intensity ratios and band profiles typical of Magellanic Cloud \HII\ regions.

Finally, we modeled the spectral energy distribution of the YSO N160\,A-IR with the dust radiative transfer model {\sc dusty}. The best model has a mixture of silicates (50\%), graphite (30\%) and amorphous carbon grains (20\%) with a single power law dependence of the density with radius proportional to $r^{-1}$, where $r$ is the radius. This corresponds to a visual magnitude of \hbox{70$\pm$10~mag} and a 9.7\,\um\, optical silicate depth of 4.3.  A bolometric luminosity of $2\times 10^5L_{\sun}$ was derived, suggesting a central star of spectral type O6\,V and about $30\,M_{\sun}$. This luminosity is about 10 times lower than the previous estimate of \cite{henning98}.

 \begin{acknowledgements}
This work is based on observations obtained at the European Southern Observatory, La Silla, Chile (ID 72.C-0600) and observations made with the \spitzer\
Space Telescope, which is operated by the Jet Propulsion Laboratory,
California Institute of Technology under a contract with NASA.
It is also based on observations with ISO, an ESA project with instruments funded by ESA Member States (especially the PI countries: France, Germany, the Netherlands and the United Kingdom) and with the participation of ISAS and NASA. 
During this work, NLMH has been supported by a Juan de la Cierva fellowship from
the Spanish Ministerio de Ciencia y Tecnolog\'{\i}a (MCyT). This work has also
been partially funded by the Spanish MCyT under project AYA2004-07466.
 \end{acknowledgements}


\end{document}